# Unique Brain Network Identification Number for Parkinson's and Healthy Individuals Using Structural MRI

**Tanmayee Samantaray [1], Utsav Gupta [1], Jitender Saini [2] and Cota Navin Gupta [1,*]**

[1] Neural Engineering Lab, Department of Biosciences and Bioengineering, Indian Institute of Technology Guwahati, Guwahati 781039, India; tanma176106113@iitg.ac.in (T.S.); ugupta@iitg.ac.in (U.G.)
[2] Department of Neuroimaging and Interventional Radiology, National Institute of Mental Health and Neurosciences, Bengaluru 560029, India; jsaini76@gmail.com
* Correspondence: cngupta@iitg.ac.in; Tel.: +91-361-2582232

**Abstract:** We propose a novel algorithm called Unique Brain Network Identification Number (UBNIN) for encoding the brain networks of individual subjects. To realize this objective, we employed structural MRI on 180 Parkinson's disease (PD) patients and 70 healthy controls (HC) from the National Institute of Mental Health and Neurosciences, India. We parcellated each subject's brain volume and constructed an individual adjacency matrix using the correlation between the gray matter volumes of every pair of regions. The unique code is derived from values representing connections for every node (i), weighted by a factor of $2^{-(i-1)}$. The numerical representation (UBNIN) was observed to be distinct for each individual brain network, which may also be applied to other neuroimaging modalities. UBNIN ranges observed for PD were 15,360 to 17,768,936,615,460,608, and HC ranges were 12,288 to 17,733,751,438,064,640. This model may be implemented as a neural signature of a person's unique brain connectivity, thereby making it useful for brainprinting applications. Additionally, we segregated the above datasets into five age cohorts: A: ≤32 years ($n1$ = 4, $n2$ = 5), B: 33–42 years ($n1$ = 18, $n2$ = 14), C: 43–52 years ($n1$ = 42, $n2$ = 23), D: 53–62 years ($n1$ = 69, $n2$ = 22), and E: ≥63 years ($n1$ = 46, $n2$ = 6), where $n1$ and $n2$ are the number of individuals in PD and HC, respectively, to study the variation in network topology over age. Sparsity was adopted as the threshold estimate to binarize each age-based correlation matrix. Connectivity metrics were obtained using Brain Connectivity toolbox (Version 2019-03-03)-based MATLAB (R2020a) functions. For each age cohort, a decreasing trend was observed in the mean clustering coefficient with increasing sparsity. Significantly different clustering coefficients were noted in PD between age-cohort B and C (sparsity: 0.63, 0.66), C and E (sparsity: 0.66, 0.69), and in HC between E and B (sparsity: 0.75 and above 0.81), E and C (sparsity above 0.78), E and D (sparsity above 0.84), and C and D (sparsity: 0.9). Our findings suggest network connectivity patterns change with age, indicating network disruption may be due to the underlying neuropathology. Varying clustering coefficients for different cohorts indicate that information transfer between neighboring nodes changes with age. This provides evidence of age-related brain shrinkage and network degeneration. We also discuss limitations and provide an open-access link to software codes and a help file for the entire study.

## 1. Introduction

Parkinson's disease (PD) is one of the most common neurological disorders that worsens with age. With a prevalence of 6.1 million people across the globe [1], it is commonly found in older people aged 50–60 [2]. Males are more likely to be affected by PD than females [3]. The associated pathological hallmark is the loss of dopaminergic neurons in the substantia nigra [4]. It leads to various classical motor and non-motor manifestations such as tremors, bradykinesia, rigidity, behavioral dysfunctions, and cognitive dysfunctions. However, the progression of neurodegeneration in PD begins and spreads throughout the nervous system long before these symptoms are expressed, termed the prodromal phase [5]. Our understanding of brain regions underlying clinical manifestations of PD is constantly growing as a result of recent developments in neuroimaging [6]. In addition to being non-invasive, structural magnetic resonance imaging (sMRI) is a robust and safe method to produce high-resolution 3D scans of the brain. PD has been associated with gray matter (GM) atrophy, detected as morphological changes by voxel-based morphometry [7,8] on sMRI. Hence, gray matter tissue contains important information to be deployed for analysis and understanding of the disease.

The effect of one neuron on another is determined by structural connectivity, which subsequently affects the potential of functional networks [9]. Thus, in diseased conditions, aberration in network topology influences the functionality of specific brain regions [10]. Structural MRI has the potential to detect alterations in the PD brain network triggered by GM volume [11,12]. Brain networks provide possible biomarkers not only for a disease cohort; however, also for individual subjects [13]. Although group-level structural network analysis has been widely explored in PD, there is minimal research on individual-level networks [14,15]. Intriguingly, individual structural connectivity analysis using sMRI may play a crucial role in analyzing brain disease [13] and brainprinting [15]. Group-level sMRI-based structural network analyses in PD have found abnormal network topology with reduced clustering coefficient, mean local efficiency [11], and lower connectivity strength [12]. Disruptive alteration was also observed in PD patients with significantly different clustering coefficients, characteristic path length, lambda and gamma. However, hemiparkinsonian PD did not show any difference in network metrics from that of controls [16]. Although individual-level network analysis has previously been conducted in other psychiatric disorders [13,17], structural connectivity studies in PD seem unexplored.

However, due to inter-individual variability, there is a demand for encoding networks into unique representations for each individual. This encoding scheme may also enable the efficient storage and transfer of individual networks. Weininger et al. [18] proposed a chemical notation specifically designed for chemical compounds, where its structure is a network, atoms are nodes, and bonds are edges. Later, Lukovits I.I. [19] obtained a compact form of a chemical compound where the presence of a bond between carbon atoms was represented as 1 in the adjacency matrix. However, this method failed for compounds with more than 15 carbon atoms, thereby demanding more investigations. This work was studied in detail and shaped further to construct UBNINs for brain networks, where brain regions are nodes and the inter-regional connections are edges.

Studies have demonstrated PD with GM atrophy; however, none have examined how Parkinson's brain networks change with aging. Age is the largest risk factor influencing the clinical progression of PD [20–22]. However, there are still undiscovered brain processes that underlie how aging affects PD neurodegeneration. A faster development of motor disability, severe gait, and postural impairment often leads to dementia in PD patients with age [23]. Moreover, PD patients have been vastly observed with an inability to understand, and process information, and other cognitive disabilities. As we age, changes in human brain structure affect not only memory and learning but also other cognitive skills. We envisage that such improper and inefficient information dissemination in PD patients may be due to aberrant brain connectivity.

In this study, our main objective is to create a unique brain identification number to represent an individual brain network. Hence, we propose a model for automatically creating a unique brain network identification number for each individual. This was inspired by an encoding technique where a chemical structure was represented as an adjacency matrix with carbon atoms as nodes and bonds as edges [19]. Additionally, we hypothesize that disease progression with age causes disruption of the brain network. Therefore, we analyzed brain network topology over a range of sparsities and evaluated connectivity metrics for different age cohorts of PD. This may address how age affects the PD brain network. These analyses were also performed on a healthy control (HC) dataset.

## 2. Materials and Methods

### *2.1. Participants*

Subjects were recruited at the Department of Neurology, National Institute of Mental Health and Neurosciences (NIMHANS), India. Every individual had provided written informed consent in compliance with the NIMHANS Institutional Ethics Committee. The data considered for our current study included 180 PD patients (135 males and 45 females;

age: 54.84 ± 9.78 years, age range: 22–72 years) and 70 HC patients (52 males and 18 females; age: 49.24 ± 10.99 years, age range: 20–73 years). The PD subject characteristics obtained were the mean Unified Parkinson's Disease Rating Scale at off-state (UPDRS Off, 33.26 ± 8.92), at on-state (UPDRS On, 17.85 ± 6.27), Hoehn and Yahr scale (H&Y, Median: 2), and Age at onset (48.27 ± 9.14). For age-based analysis, the participants were categorized into five age groups (age group A: ≤32 years (n1 = 4, n2 = 5), B: 33–42 years (n1 = 18, n2 = 14), C: 43–52 years (n1 = 42, n2 = 23), D: 53–62 years (n1 = 69, n2 = 22) and E: ≥63 years (n1 = 46, n2 = 6), where n1 and n2 are the number of subjects in PD and HC, respectively. Brain parcellation for one of the PD subjects in age group D was unsuccessful, due to which it was excluded from the study. The age-based participants' demographics are listed in Table 1. Our study uses a retrospective dataset that has been previously employed on PD [11,24].

**Table 1.** Demographic and clinical information of participants.

| Age-Cohort | Count | | Gender (Male:Female) | | UPDRS off | UPDRS on | H&Y | Age at Onset |
|---|---|---|---|---|---|---|---|---|
| (Age Range) | PD | HC | PD | HC | (Mean ± SD) | (Mean ± SD) | (Median) | (Mean ± SD) |
| A (≤32) | 4 | 5 | 3:1 | 5:0 | 31.69 ± 2.71 | 16.75 ± 9.91 | 2 | 16.83 ± 10.67 |
| B (33–42) | 18 | 14 | 11:7 | 7:7 | 32.94 ± 7.13 | 16.67 ± 6.16 | 2 | 35.19 ± 4.69 |
| C (43–52) | 42 | 23 | 28:14 | 19:4 | 32.33 ± 11.11 | 17.78 ± 8.15 | 2 | 42.83 ± 4.29 |
| D (53–62) | 69 | 22 | 56:13 | 16:6 | 32.79 ± 8.76 | 17.70 ± 5.37 | 2 | 47.47 ± 6.26 |
| E (≥63) | 46 | 6 | 36:10 | 5:1 | 35.17 ± 6.97 | 19.30 ± 4.78 | 2 | 51.54 ± 9.88 |
| *p*-value | | | | | 0.58 | 0.54 | | $3.08 \times 10^{-22}$ ** |

A, B, C, D, E: Age-based groups with age range in brackets; H&Y: Hoehn and Yahr scale; *p*-value: from ANOVA test; SD: Standard Deviation; UPDRS: Unified Parkinson's Disease Rating Scale; ** Significant at 99% confidence interval; H&Y (categorical value) is expressed as median.

### *2.2. MRI Scanning and Image Processing*

A 3 Tesla Philips Achieva scanner with a 32-channel head coil was used to scan the participants. A high-resolution T1-weighted structural MRI scan was obtained with a magnetization-prepared rapid acquisition gradient echo (MPRAGE) sequence (repetition time (TR) = 8.2 ms, echo time (TE) = 3.7 ms, flip angle = 8, field of view (FOV) = 256 × 256 × 165 $mm^3$, 165 sagittal slices, voxel size = 1 × 1 × 1 $mm^3$). The original scans were obtained in DICOM format, which were converted into NIfTI by using MRIcron's dcm2nii tool (http://www.nitrc.org/projects/mricron (accessed on 7 May 2018)). The pre-processing steps are explained elsewhere [11]. This involved the MATLAB-based Computational Anatomy Toolbox (CAT12, (vCAT12.8), http://www.neuro.uni-jena.de/cat/) (accessed on 7 May 2018) within Statistical Parametric Mapping (SPM12, v7771). A recent study [25] has inferred better age prediction using this pipeline, and hence we used it for brain network analysis. The scans were segmented and smoothed before taking the GM map for further study.

### *2.3. Individual Brain Network Construction*

The GM map of every individual was parcellated into 56 regions using LPBA40, the LONI Probabilistic Brain Atlas [26] (https://resource.loni.usc.edu/resources/atlases-downloads/ (accessed on 7 May 2018)). The regions are illustrated in the Appendix of [11]. A linear regression analysis was performed to remove the effect of gender. Each brain region indicated a node in the network for which region-specific gray matter volumes (rGMV) were obtained using CAT12. Individual brain networks were constructed [13] for each subject based on correlation $(r_{ij})$ between the rGMV of each pair of regions. It was computed using Equation (1), where $rGMV_i$ and $rGMV_j$ are the regional GM volumes of regions $i$ and $j$.

$$r_{ij} = \frac{1}{\left(rGMV_i - rGMV_j\right)^2 + 1} \quad (1)$$

The correlation or association matrix, hence obtained, was visualized as a network in BrainNet Viewer (Version 1.7) [27], with coefficients representing edges and regions indicating nodes.

### *2.4. Generating Unique Brain Network Identification Number (UBNIN)*

Our algorithm uses the binary adjacency matrix (AM) of a network (N). Every individual association matrix was binarized using consistency-based thresholding to retain 30% of connections [28,29]. Given that AM is symmetric, it suffices to consider either the upper or lower triangle of AM. Let us consider a sample network (Figure 1) with 10 nodes whose upper triangle matrix is AM (Figure 2). The lower triangular matrix (shaded in blue in Figure 2, step 1) is zeroed due to the symmetric matrix and undirected network, hence not included in computing UBNIN. The entry of any column is either 0 or 1, and the entire column is assumed to be a single binary number starting from bottom to top (step 2, Figure 2).

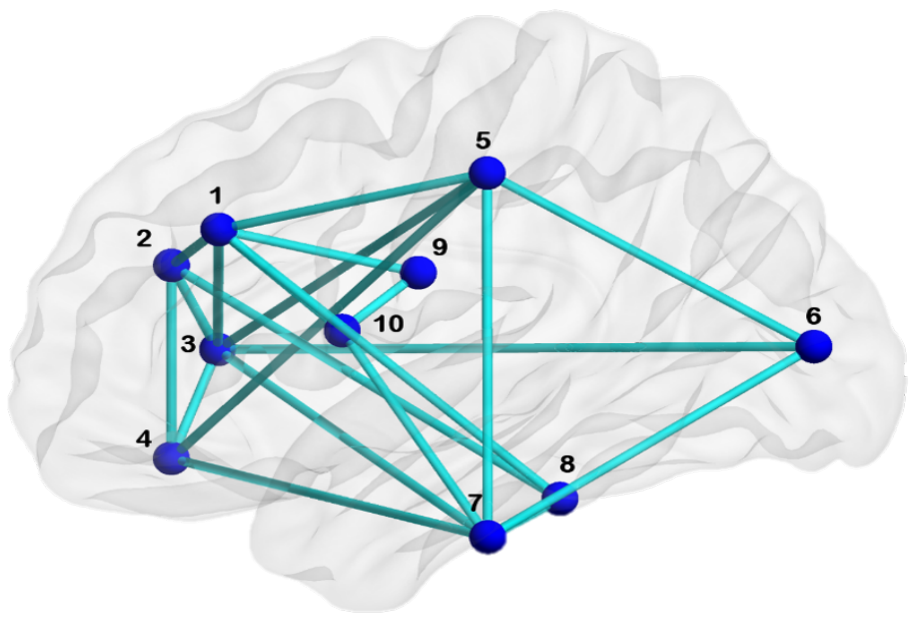

**Figure 1.** Sample network for deriving UBNIN.

So, every column of AM above the principal diagonal was transformed into a decimal equivalent (step 3, Figure 2) [19]. Hence, the decimals thus obtained were 1, 3, 6, 13, 20, 60, 67, and 1, 321 to be stored in an array DEC (i.e., decimal equivalents). It is apparent that the minimal decimal value of any column is zero; the maximal value of a column *i* is $2^{i-1}$, where *i* = 1 for the first column of the upper triangle of AM. A node with a zero decimal value indicates a disconnected node, while a decimal value of $2^{i-1}$ indicates a node connecting to every other node. The maximal decimal equivalents of the columns (nodes) are 1, 3, 7, 15, 31… $2^{n-1}$ for a fully connected network with n nodes. The values in DEC were further converted into a single numerical form, named $UBNIN_T$, using Equation (2), where T is the total number of nodes in the network or the number of columns in the binary matrix.

$$UBNIN_{T=10} = \frac{1}{2^{i=8}} * \left[\!\left[\frac{1}{2^{i=7}} * \left\{\frac{1}{2^{i=6}} * \left\|\frac{1}{2^{i=5}} * \left(\frac{1}{2^{i=4}} * \left[\frac{1}{2^{i=3}} * \left\langle\frac{1}{2^{i=2}} * \left\lfloor\frac{1}{2^{i=1}} * DEC(2) + DEC(3)\right\rfloor + DEC(4)\right\rangle + DEC(5)\right] + DEC(6)\right) + DEC(7)\right\| + DEC(8)\right\} + DEC(9)\right]\!\right] + DEC(10) \quad (2)$$

Equation (2) upon substituting corresponding decimal and node values from Figure 2 gives:

$$UBNIN_{T=10} = \frac{1}{2^8} * \left[\!\left[\frac{1}{2^7} * \left\{\frac{1}{2^6} * \left\|\frac{1}{2^5} * \left(\frac{1}{2^4} * \left[\frac{1}{2^3} * \left\langle\frac{1}{2^2} * \left\lfloor\frac{1}{2^1} * 1 + 3\right\rfloor + 6\right\rangle + 13\right] + 20\right) + 60\right\| + 67\right\} + 1\right]\!\right] + 321$$

This results in $UBNIN_{T=10} =$ 321.005979848895, for network (Figure 1) with 10 nodes.

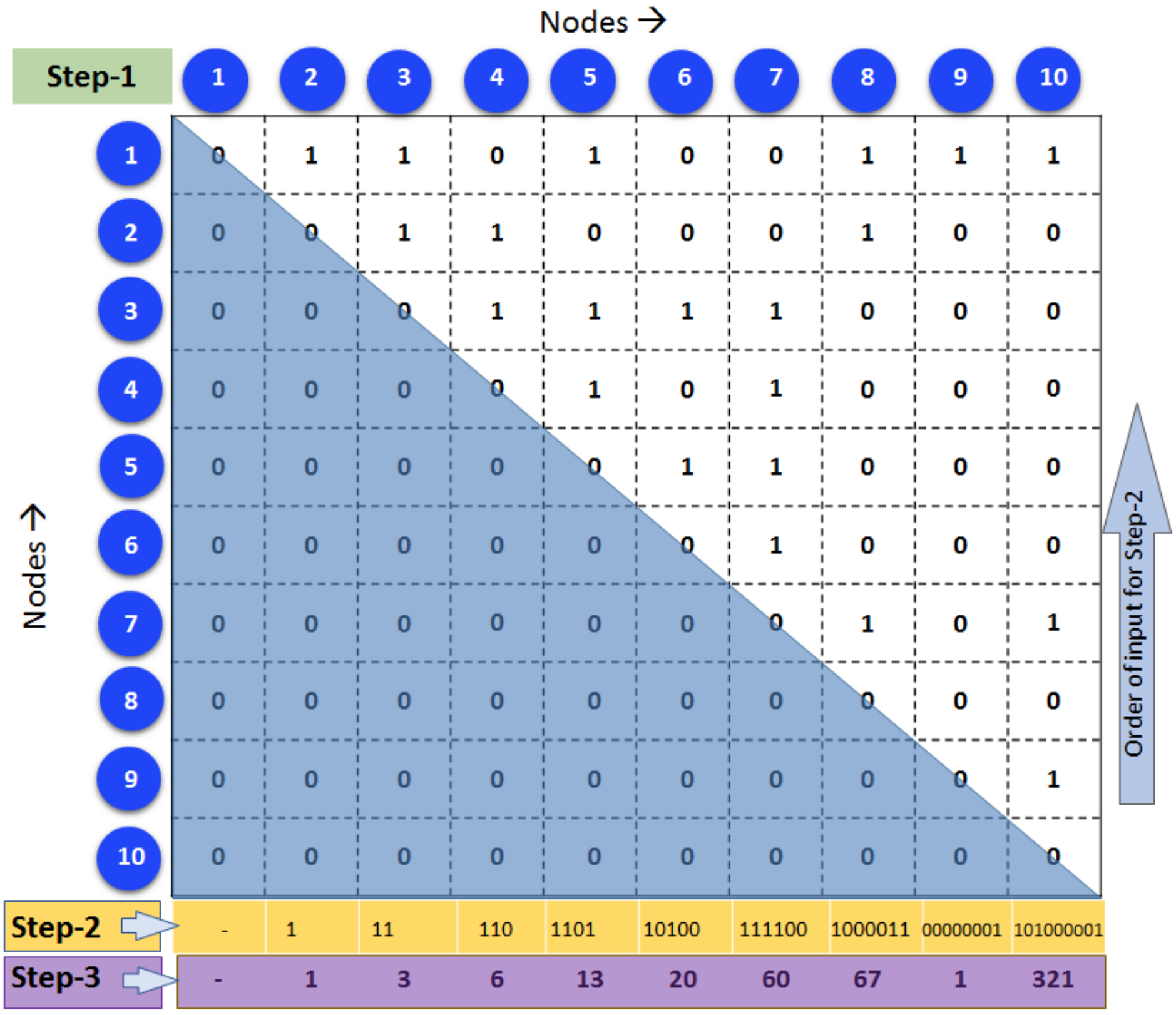

**Figure 2.** Adjacency matrix corresponding to sample network (as in Figure 1). Step 2 and Step 3 are intermediate steps.

The power of $\frac{1}{2^i}$ starts with $i = 1$ and the equivalent decimal (*DEC*) starts with the second node of the adjacency matrix. The top-to-bottom approach for this methodology was also attempted for multiple adjacency matrices; however, the UBNIN thus obtained was larger than that from bottom-to-top. Hence, the binary number (bottom-to-top) of each node was deployed to convert into its corresponding decimal equivalent and applied in (1) to obtain the UBNIN. The developed algorithm is illustrated as Algorithm 1.

**Algorithm 1:** UBNIN for an Adjacency matrix (AM) of a given network.

```
N = length(AM); AM = upptriang(AM);
for j = 2:N
    BinNode = AM(j − 1:(−1):1, j);
    for a = 1:length(BinNode)
        Val+ = 10^(j−a−1)(BinNode(a));
    end
    DEC_j = decimal(BinNodeVal);
end
temp = DEC_2;
power = 1;
for i = 2:N−1
    UBNIN_T = temp*(1/2^power);
    UBNIN_T+ = DEC_i+1;
    power = power+1;
end
```

*2.5. Age-Based Network Metric Analysis*

A subject-by-regions matrix for each age group was used as input, where subject count varied with age cohort while the number of regions remained constant. A weighted undirected matrix (association matrix) was constructed [11] using Pearson's correlation between the regional GM volume of every pair of nodes in MATLAB (MathWorks R2020a). Pearson's correlation gives the measure of association between brain regions (nodes) and is used if the number of subjects is less than the number of regions [8,16,30].

To facilitate simplified network analysis, the association matrix of each age cohort was binarized across an optimized range of sparsities. Sparsity implies the densities of zeros in a matrix and is used as a measure of threshold, yielding an adjacency matrix by removing spurious edges. A sparsity threshold of 0.8 considers the highest 80% values only and rejects the lowest 20% values. The range is optimized in such a manner that below the lower limit, the network becomes equivalent to a random network (i.e., small-world index close to 1). And above the upper limit, the average degree becomes less than the logarithm of the number of nodes, and the small world index is not estimable [11,31,32].

A complex brain network is often measured and analyzed in terms of its clustering coefficient. The nodal clustering coefficient ($C_i$) is defined as the fraction of directly linked neighbors that are interconnected among themselves [32], given as Equation (3).

$$C_i = \frac{2t_i}{k_i * (k_i - 1)} \quad (3)$$

where $i$ is node; $k_i$ is the total number of nodes; $t_i$ is total number of edges between neighbors of $i$. The nodal clustering coefficients were estimated using Brain Connectivity Toolbox-derived MATLAB functions [32] and averaged over the network to obtain the mean clustering coefficient, separately performed for HC and PD. A structural brain network is a pattern containing edges between two nodes where interregional correlation exists. The networks for each age cohort were visualized using BrainNet Viewer (Version 1.7) [27].

### *2.6. Statistical Analysis*

An analysis of variance (ANOVA) was performed between PD age groups to find differences in clinical variables such as UPDRS off, on, and age at onset. A $p$-value < 0.05 was concluded to be significant. Inter-group differences in clustering coefficient were examined in both HC and PD using a nonparametric permutation test [11,33] with 1000 iterations to determine their statistical significance. While preserving the number of subjects in each group, clustering coefficients were randomly assigned to either group. Randomized group differences determined the permutation distribution of the differences. Thus, actual inter-group differences were compared using self-developed MATLAB code.

## 3. Results

### *3.1. UBNIN for Individual Brain Network*

In this study, we proposed a unique brain network identification number (UBNIN). It is a numerical representation scheme that was observed to be distinct for every individual network. The weighted undirected matrix, hence formed, contained 56 nodes and 1540 (= 56 × 55/2) edges. Adjacency matrix for each patient was constructed using 30% consistency-based thresholding. The UBNIN of PD and HC individuals, computed using the given MATLAB-based $p$ code, were all distinct and ranged between 15,360 to 17,768,936,615,460,608 and 12,288 to 17,733,751,438,064,640, respectively.

### *3.2. Age-Based Network Metric Analysis*

A significant difference was found in age at onset ($p = 3.08 \times 10^{-22}$) between the age-based groups of PD at the 99% confidence interval (as in Table 1). However, no significant intergroup difference was observed in UPDRS at the Off ($p = 0.58$) and On ($p = 0.54$) states (Table 1 and Figure 3). Detailed demographic information for various age groups of HC and PD is in Table 1, and group comparisons of clinical features of PD are reported in Table 1 and Figure 3. Pearson correlation was performed to detect relationships between regions. The weighted undirected correlation matrix for PD is shown in the top row of Figure 4. The strength of the connections is indicated by the color bar.

The binary undirected correlation matrix of PD age groups is shown in the middle row of Figure 4. The axial view of undirected networks for each PD age group is shown in the bottom row of Figure 4, where intra- and inter-hemispheric connections can be seen. These are shown at the same sparsity threshold of 0.8 for illustration purposes only.

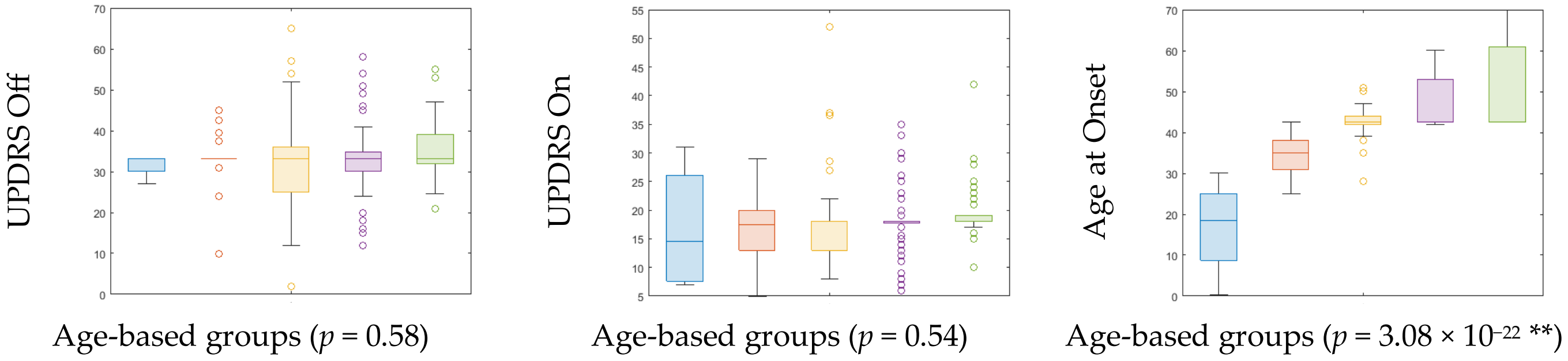

**Figure 3.** Plotted are ANOVAs of PD clinical features across age groups. UPDRS: Unified Parkinson's Disease Rating Scale; Notches indicate median; Top, bottom edges of the box indicate 25th and 75th percentiles; ** Significant at 99% confidence interval. Blue color indicates age-group A, Orange color indicates age-group B, Yellow color indicates age-group C, Violet color indicates age-group D, and Green color indicates age-group E.

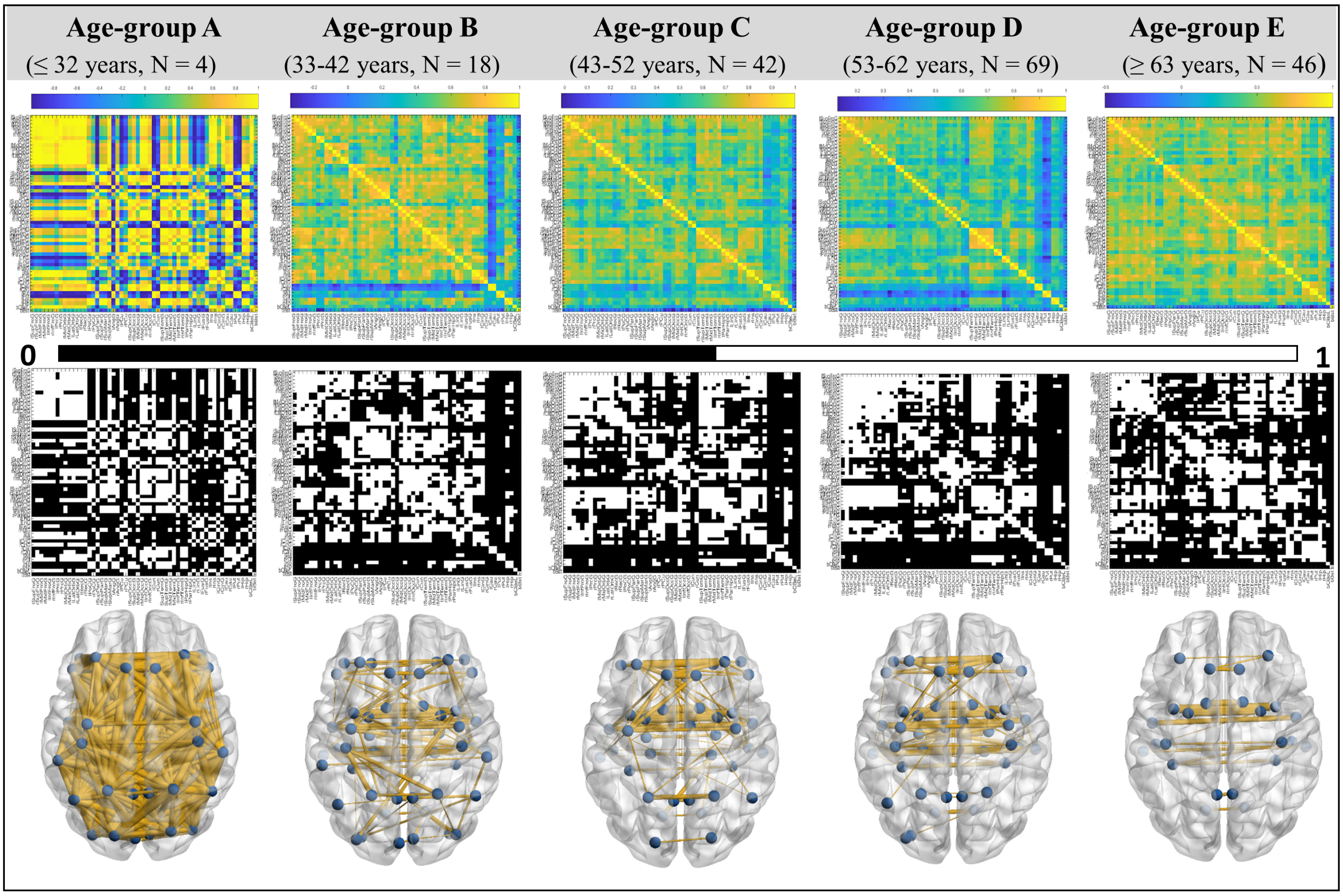

**Figure 4.** Age-based Weighted Undirected Matrix (top row), Binary Undirected Matrix (middle row), and Undirected Network (bottom row) of PD. Color-bar shows the strength of connections. N is the number of subjects in a specified age range. Networks are at the 0.8 threshold for illustration purposes only. The *x*- & *y*-axes in weighted and binary undirected matrices are brain regions in order as listed in [13,22].

The sparsity threshold was estimated from the data, and every age cohort was analyzed in a range of 0.6 to 0.9 with a step size of 0.03 [11,31]. The small-world index exceeded 1 for every threshold level. The clustering coefficient of the networks obtained was higher, and at the same time, the mean shortest path length was equivalent to that of a random network in the sparsity range. Figure 5 illustrates the distribution of the mean clustering coefficient and the significant comparison of different age groups as obtained from the adjacency matrix. The mean clustering coefficient in PD declined over sparsity except for group A at 0.87 sparsity (Figure 5i), suggesting cognitive decline with aging in PD patients. However, more investigation is encouraged on a large sample of young PD patients to confirm the findings in group A. Interestingly, Figure 5vi illustrates a reduction in the mean clustering coefficient in HC with age, except for some variability in groups B (at 0.78 sparsity) and E (at 0.72 sparsity). This decline in network metrics indicates a change in network topology, which may be due to natural aging in HC. The blue circles (Figure 5ii–v) are mean differences obtained from averaging 1000 permutations. The mean clustering coefficient of group A was observed to be significantly higher than that of the B, C, D, and E age groups. The obvious fact that there are fewer patients at younger ages, especially those younger than 32 years, calls for further research on a larger dataset.

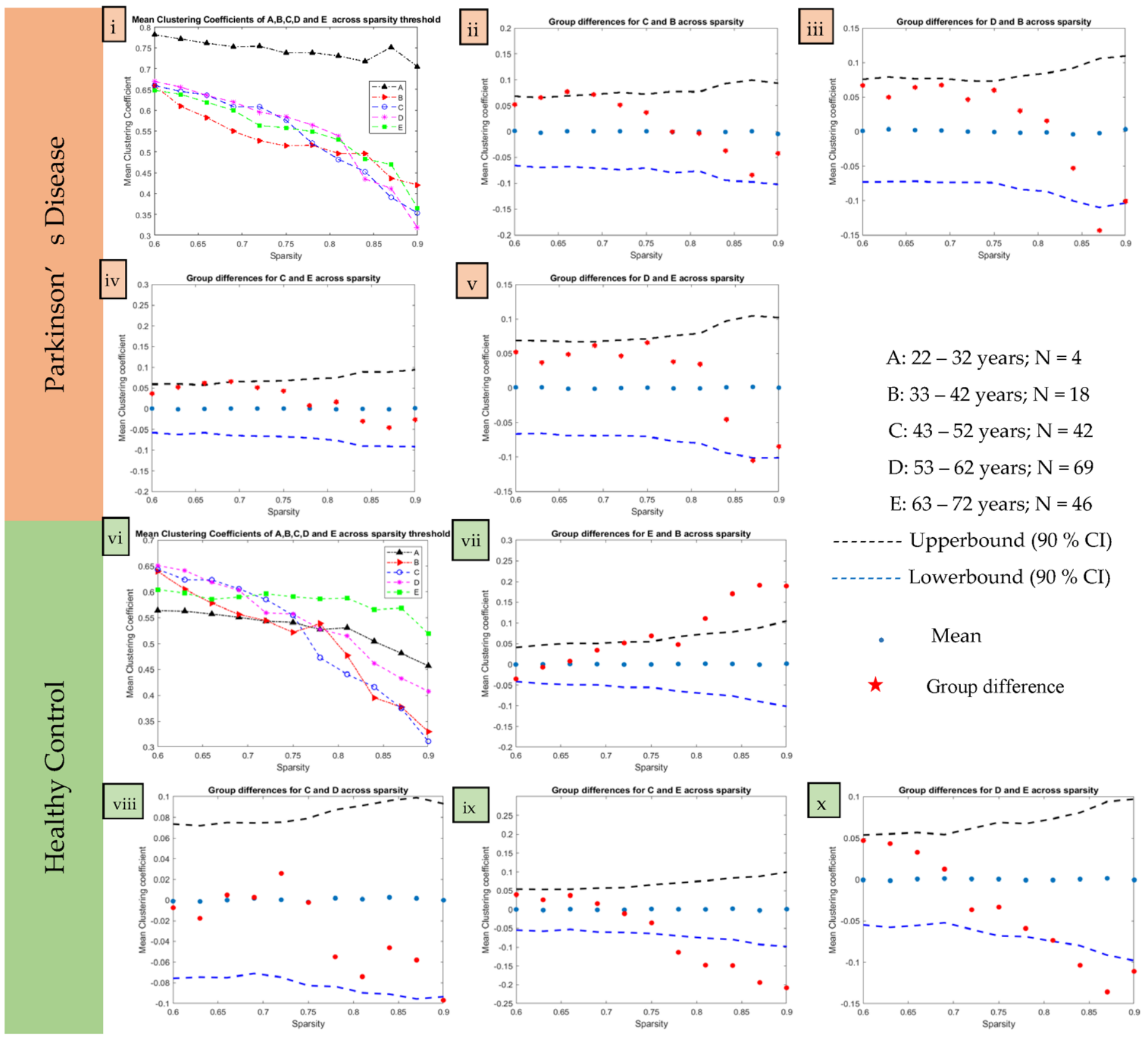

**Figure 5.** Distribution and intergroup differences in mean clustering coefficient for age-based groups. The subfigures are numbered ((**i**–**v**) for PD and (**vi**–**x**) for HC) and colored as orange for PD and green for HC individuals. The red dots indicate differences between age-groups in Healthy controls.

In PD, the clustering coefficient of group C (between 43 and 52 years) is significantly different from that of PD group B (33–42 years) at the 0.66 sparsity level (Figure 5ii). The clustering coefficient of PD group C (between 43 and 52 years) is significantly different from that of PD group E (≥63 years) at 0.66 and 0.69 sparsities (Figure 5iv). It is overall higher in PD group C than PD group E at lower sparsities up to 0.75; however, lower at higher sparsities from 0.84 onwards. Moreover, the clustering coefficient of PD group D (53–62 years) is significantly different from that of PD group B (33–42 years) at higher sparsity levels of 0.87 and 0.9 (Figure 5iii). It is higher in PD group D than B until 0.81 sparsity; however, afterward, it is lower. The clustering coefficient in PD group D is significantly different from that of PD group E (≥63 years) at 0.87 sparsity (Figure 5v). PD group D has a larger clustering coefficient than PD group E up to 0.81 sparsity; however, after that it decreases. However, other group comparisons showed no difference between PD group B (33–42 years) with PD group E (≥63 years) and PD group C (43–52 years) with PD group D (53–62 years).

A similar permutation analysis in HC showed the clustering coefficient for age group A was significantly different from every other group. However, HC group E (≥63 years) was significantly different from that of HC group B (33–42 years) at sparsities of 0.75 and above 0.81 (Figure 5vii), HC group C (43–52 years) at sparsities above 0.78 (Figure 5ix), and HC group D (53–62 years) at sparsities above 0.84 (Figure 5x). The clustering coefficient of HC group C also differed significantly from that of HC group D at a sparsity of 0.9 only (Figure 5viii). However, the same analysis may be replicated on a larger HC dataset, especially in age groups A and E.

## 4. Discussions

### *4.1. UBNIN for Individual Brain Network*

Our proposed algorithm (UBNIN) enables the numerical representation of individual brain networks and, hence, holds significance in encoding. A similar approach named BIN [19] was applied to a network where carbon atoms were nodes and bonds were edges. Although it could compress the adjacency matrix, it increased exponentially with the number of nodes and hence had limited applicability in chemical structures with only 15 nodes. However, our method can compute the encoded numerical representation for a fully connected network with up to 1024 nodes, after which it generates not a number (*NaN*) in MATLAB (R2020a) (Appendix A). The method described in [19] generates an infinitely large value (*Inf*) even for a small network with 46 nodes, which is uninterpretable. For networks containing more than 1024 nodes, we suggest using ,to use $\frac{DEC(i)}{2}$, i.e., half of the equivalent decimals, instead of direct decimals in our equation. Previous studies have confirmed that structural connections in the brain are unique to each individual person [14,15]. Moreover, the architecture of region-to-region connectivity reflects the complex wiring system or connectome of the individual which has been used for brain fingerprinting [15,34]. Hence, the numerical representation system proposed in our study has the potential to be used as a neural signature of a person's unique structural connectivity. The future of individualized mental health care may lie in this brain fingerprinting, which has demonstrated promising markers for mental illness [35]. Researchers may be able to separate various neuropsychiatric illnesses using connectivity patterns and detect network-level abnormalities in specific patients. In the end, such information might enable an important change in clinical neuroscience and a personalized prognosis for individual treatment [36]. Another aspect that could be explored is the dynamic behavior of the human brain as a result of brain plasticity. Degeneration and compensatory mechanisms play a vital role in the human brain, leading to reorganization of brain structure, functions, and connections. Structural MRI helps quantify spatial patterns, is more stable during data acquisition compared to other neuro-imaging modalities, and does not depend on the physical condition of the subject. The results reported in the current study are based on implementing our algorithm using sMRI. Each time, the algorithm produces the same UBNIN value on any given network and, hence, is representative of the network. However, any change in MRI scans leads to alterations in the brain network, which will ultimately be reflected by a different UBNIN value. The same may be tested on longitudinal data of PD patients to analyze the network change over time by tracking changes in UBNIN values. We further tested it on healthy controls and observed that every individual had a unique UBNIN. However, the individuals cannot be directly categorized based on their UBNIN values because of different network topologies and connection patterns. A higher UBNIN does not necessarily imply a denser network; it depends on the nodes to which it is connected. Additionally, we have also implemented the algorithm on multiple (count = 100,000) random binary matrices of dimension 10 × 10 and found distinct UBNIN values each time.

Threshold value affects the adjacency matrix for computing individual UBNIN from structural MRI, and hence it is crucial to select an appropriate thresholding method. UBNIN is also advantageous in saving storage memory and offering low-space, high-speed transferring of network information. The adjacency matrix of an individual is always a 2-D symmetric matrix, and the corresponding network consumes a lot of computer storage space and takes time and space to transfer. The number of rows and columns in the adjacency matrix, however, depends on the brain atlas used during parcellation. Here in our case, it is 56 × 56, which may increase to a higher count in other brain atlases. Moreover, if a sparsity thresholding measure is applied (e.g., 10 different sparsities) for 1000 subjects, we obtain 10,000 numbers of 2D matrices. This imposes a storage burden on computer systems and databases. The idea emerged while performing network analysis on multiple adjacency matrices for the age cohorts. Not only vast computer memory but also huge computation time were consumed, which motivated us to find an encoding scheme for these networks. Hence, our encoding approach (UBNIN) may help overcome the problem and be applied to any network with a large number of nodes.

### *4.2. Age-Based Network Metric Analysis*

Although the standard parcellation method has been employed in our study, the use of connectivity-based parcellation is encouraged in PD [37]. The results (from the 4th and 5th columns of Figure 4) accentuate the progressive decline in structural connectivity with age, which indirectly affects the functional characteristics, which corroborates the earlier findings [38]The brain network states the pattern of connections between neuronal elements and explains information processing inside the brain [8,11]. Connectivity studies using sMRI have analyzed brain networks in PD compared to healthy [11,16,33]controls , across gender [39], and with different associated dysfunctions such as tremor [40] and mild cognitive impairment [12]. Hence, there is a lack of structural brain connectivity studies in PD across the lifespan.

We obtained a significantly higher mean clustering coefficient in group A than in every other age cohort. This suggests higher information integration in young adults, i.e., up to 32 years old. This is in agreement with a previous study [41] on healthy individuals based on electroencephalogram (EEG) signals. The functional connectivity network from the beta-band working memory task evoked a significant increase in clustering coefficient for younger adults (19–29 years) compared to that of older participants (58–70 years) [41]. However, due to the limited subjects in group A, we suggest additional investigation with a larger sample size. It is intriguing to note that inter-group differences were observed in almost all PD age groups: B (33–42 years) vs. C (43–52 years), B vs. D (53–62 years), C vs. E (≥63), D vs. E. This may be suggestive of the fact that the clustering coefficient in PD varies with age, affecting the network organization and its ability to aggregate information from different nodes. Hence, the disease affects different age groups differently. Network degeneration is also consistently found in healthy people due to natural aging. The differences in clustering coefficient are evident in almost all pairs of HC age groups (except for B vs. C and B vs. D). This research may be further replicated on other large datasets. Since the clustering coefficient measures functional segregation [42], functional connectivity studies have shown it decreasing with age [43], which corroborates with our result. Network metrics have shown high correlation with each other [28], which may be investigated in age groups of PD patients from multiple sites and scanners. Due to the limited availability of clinical features in our study, more research may confirm the clinical changes with age. However, previous studies [22] have confirmed that age strongly affects the clinical manifestation of PD. Additionally, functional analysis may provide a complementary understanding of functional changes taking place along with changes in network topology.

The primary risk factor associated with PD, which also contributes to its onset and progression, is aging. Our results indicate that chronological age is proportional to the age at disease onset. However, the age of a patient rather than age at onset is a factor that greatly affects PD progression, including the emergence of dementia and hallucinations [44]. The clinical manifestation of PD may change as people age, and these changes may include changing pharmacological side effects, an increased risk of dementia, hallucinations, and an increased likelihood of nursing home admission. Hence, age-related brain changes in PD are reflected by motor and non-motor dysfunctions as a phenotypic presentation of the disease [44]. The neuronal processes underlying the consequences of aging on the neurodegeneration of Parkinson's disease, however, remain unanswered. In the present research, we discovered that aging significantly affects the gray matter networks of Parkinson's patients. This is evident from the constant fall in clustering coefficient with sparsity in age groups, with a slight variability that may be due to compensatory [40] mechanisms . As gray matter covers the neuronal cell, we assume neuronal decay over time affects the gray matter volume and thereby leads to network aberration. This may offer a potential explanation for how aging leads to neurodegeneration in PD due to the fact that information is perceived differently. Results from a diffusion tensor imaging study showed Parkinson's patients at older ages exhibited disrupted white matter network topology compared to young age groups [45]. The clustering coefficient of PD patients aged 69–82 years was found to be lower than that of patients aged 34–69 years. However, the same study did not find any difference in gray matter covariance networks using similarity-based methods on structural MRI data or functional networks over age [45]. Age-associated structural brain networks were analyzed in order to better comprehend the underlying connections across ages. It may assist researchers and physicians in offering age-specific treatments. Anti-aging medicines might open new doors for the creation of disease-modifying therapeutics for PD [46,47] to interfere with the pathophysiology of the disorder [48,49]. We assume our investigation may play a new role in assisting researchers to understand variation in connections across ages.

### *4.3. Limitations and Future Scope*

The entirety of the UBNIN value can be precisely calculated manually (using paper and pen) as a unique identifier for a binary matrix, as can the reconstruction of the original binary matrix. The current computer hardware technology limits the number of significant digits, which slightly hampers the accurate reconstruction of larger binary matrices.

For a larger 20 × 20 fully connected binary matrix, 190 digits would be required to represent the collection of decimal values. The UBNIN values hence obtained are rounded up (as shown in the Appendix), which is a computational constraint. This limitation is illustrated in the excel file (sheet: Demonstrating Precision Issue), where the multiplication factor ($\frac{1}{2^i}$) rounds up at large i (i.e., node) values.

The network pattern changes over time [50], with varying degrees of change, both in normal healthy [15,41,51] controls and different disease conditions . Although previous studies have explored this phenomenon using diffusion MRI [14,15] or electroencephalography [34] structural MRI data, due to its lower temporal resolution, may give a more stable result. In addition, individual brain networks from sMRI, UBNIN may be applied to other imaging modalities as well. It may also be implemented to encode any binary network, viz., the internet, citation networks, economic networks, and even social networks. Earlier research [15] has observed that the brain's unique pattern changes over time and is also affected by disease, the environment, and other factors. This UBNIN, verified using structural MRI, may pave the way towards understanding brain plasticity. The individual subject's UBNIN may be further investigated to find an association with IQ as well as other clinical features of PD.

Since prodromal stage is a crucial time in PD and is not explored yet, more research is required in this direction. Moreover, longitudinal study may help understand the underlying change in brain network over time due to the disease. The neurodegeneration in the disease may thus be understood clearly in PD. Although PD is uncommon at younger age and the immensely dense connectivity at younger age (≤32 years) is an obvious truth, the connectivity could be investigated in detail with large sample size of younger patients to better understand the effect of Parkinson's disease. The HC individuals were also limited and hence the analysis may be replicated on a larger dataset. Employing connectivity information from other imaging modalities may help draw more firm conclusions about aging in Parkinson's disease as evident from loss of connections and differences in network metric.

## 5. Conclusions

We proposed a novel encoding system, namely the Unique Brain Network Identification Number (UBNIN), for individual brain networks. To our knowledge, this is the first-of-its-kind study on Parkinson's disease employing sMRI information. Moreover, the numerical representation of the brain network could facilitate efficient transmission and storage of network information. A change in the encoded number might reflect a change in the network pattern. Since every individual has a unique pattern of brain networks, our approach may be employed for creating individual brain IDs to store in a database and also for further analysis. It also has potential for brainprinting as it stores the unique identity of the person. and could be taken up as a research prospect for understanding brain plasticity. We have also investigated the age-based network topology and performed connectivity analysis. Our investigation of an sMRI-based structural network demonstrated varying clustering coefficients at different ages in PD. As the clustering coefficient reveals statistically significant group differences, it could be treated as a biomarker to understand information communication at different ages. Our study could be taken as a base to analyze other connectivity metrics so as to perform a comprehensive connectivity analysis.

**Author Contributions:** The inception of the idea and revision of this paper were conducted by T.S., J.S. and C.N.G. T.S. proposed the idea to C.N.G. T.S. and U.G. built up the UBNIN scheme. C.N.G. and T.S. wrote the paper. J.S. contributed to revising the manuscript. Portions of this work have been submitted by U.G. and T.S. for award of B. Tech. and Ph. D. degrees at the Indian Institute of Technology, Guwahati. All authors have read and agreed to the published version of the manuscript.

**Funding:** T.S. was funded by a Ministry of Education (MoE) doctoral scholarship by the Government of India. C.N.G.s time was supported by the Scheme for Promotion of Academic and Research Collaboration (SPARC Grant), Government of India, Project Code: P1073.

**Institutional Review Board Statement:** The study was conducted and approved by the Institute Ethics Committee of National Institute of the Mental Health and Neurosciences, India.

**Informed Consent Statement:** Written informed consent from every individual was obtained in compliance with the National Institute of Mental Health and Neurosciences (NIMHANS, India) Institutional Ethics Committee.

**Data Availability Statement:** The imaging and clinical data were collected from the National Institute of Mental Health and Neurosciences and may be provided upon approval as per Institute norms. The code developed for this work is available at our below GitHub link: https://github.com/NeuralLabIITGuwahati/UBNIN. We also provide the reconstruction of the binary matrix (Figure 2, step 1) for the sample network (Figure 1) in the excel file uploaded in the above GitHub link.

**Conflicts of Interest:** We declare that we have no known conflict of interest that could be viewed as influencing the work reported in this paper.

## Appendix A

| Number of Nodes | UBNIN Value |
|---|---|
| 10 | 511.999999999985448084771633148193359375 |
| 20 | 524288 |
| 30 | 536870912 |
| 40 | 549755813888 |
| 50 | 562949953421312 |
| 100 | 633825300114114700748351602689 |
| 150 | $7.13623846352979940529142984724747568191 \times 10^{44}$ |
| 200 | $8.03469022129495137770981046170581301261 \times 10^{59}$ |
| 250 | $9.04625697166532776746648320380374280104 \times 10^{74}$ |
| 300 | $1.01851798816724304313422284420468908053 \times 10^{90}$ |
| 500 | $1.63669530394807093500659484841379957611 \times 10^{150}$ |
| 800 | $3.33400721643992713703992589536062889857 \times 10^{240}$ |
| 1000 | $5.35754303593133660474212524530000905281 \times 10^{300}$ |
| 1020 | $5.61779104644473721165407872121570229256 \times 10^{306}$ |
| 1024 | $8.98846567431157953864652595394512366809 \times 10^{307}$ |
| 1025 | NaN |